\documentclass[12pt]{elsart4}

\usepackage{graphicx,amsmath,amssymb}

\begin{document}

\begin{frontmatter}

\title{Time Reversal Mirror and Perfect Inverse Filter in a Microscopic
Model for Sound Propagation}

\author{Hern\'an L. Calvo},
\author{Ernesto P. Danieli},
\author{Horacio M. Pastawski\corauthref{cor}},
\corauth[cor]{Corresponding author.}
\ead{horacio@famaf.unc.edu.ar}

\address{Facultad de Matem\'atica, Astronom\'ia y F\'isica, Universidad
Nacional de C\'ordoba, Ciudad Universitaria, 5000 C\'{o}rdoba, Argentina}

\begin{abstract}
Time reversal of quantum dynamics can be achieved by a global change of the
Hamiltonian sign (a hasty Loschmidt daemon), as in the Loschmidt Echo 
experiments in NMR, or by a local but persistent procedure (a stubborn
daemon) as in the Time Reversal Mirror (TRM) used in ultrasound acoustics.
While the first is limited by chaos and disorder, the last procedure seems
to benefit from it. As a first step to quantify such stability we develop a
procedure, the Perfect Inverse Filter (PIF), that accounts for memory
effects, and we apply it to a system of coupled oscillators. In order to
ensure a many-body dynamics numerically intrinsically reversible, we
develop an algorithm, the pair partitioning, based on the Trotter strategy
used for quantum dynamics. We analyze situations where the PIF gives
substantial improvements over the TRM.
\end{abstract}
\end{frontmatter}
\begin{figure}
\begin{center}
\includegraphics*[width=7.5cm]{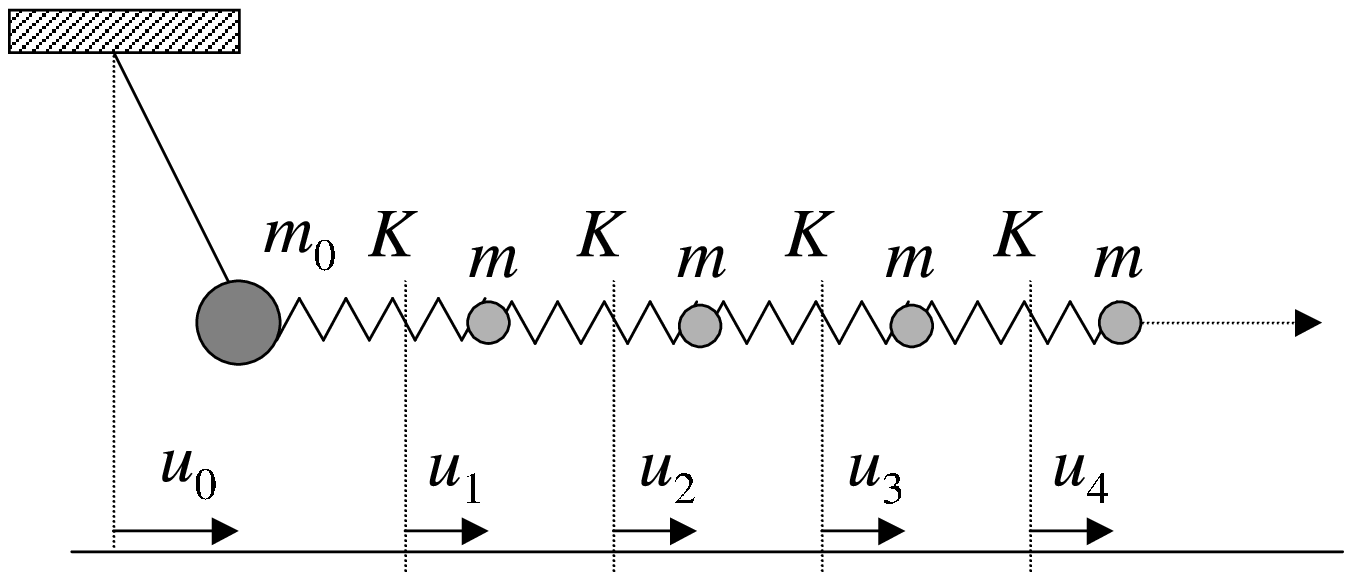}
\end{center}
\caption{Scheme of the model: a simple pendulum (surface oscillator) is
coupled to the bulk masses.}
\label{fig_scheme}
\end{figure}
\begin{figure}
\begin{center}
\includegraphics*[width=7.5cm]{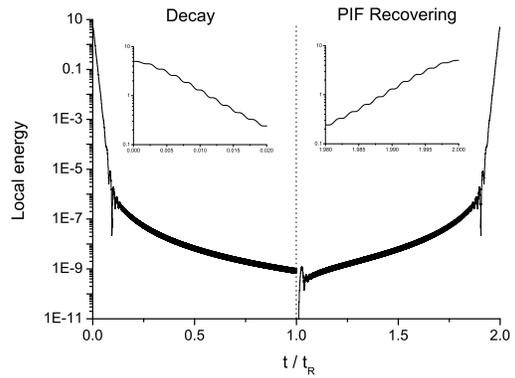}
\end{center}
\caption{Decay and recovering by PIF procedure of the local energy in
logarithmic scale. The insets show in detail the quadratic and exponential
regimes.}
\label{fig_decrec}
\end{figure}
\begin{figure}
\begin{center}
\includegraphics*[width=7.5cm]{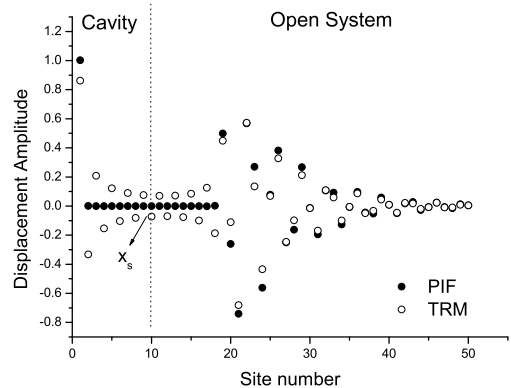}
\end{center}
\caption{Recovering of the initial condition for the two procedures.}
\label{fig_recovering}
\end{figure}

In the last years, the group of M. Fink in Paris developed an experimental
technique called Time Reversal Mirror (TRM) that allows the time reversal of
acoustic excitations \cite{Fink}. An ultrasonic pulse, produced inside a
control region (also called cavity) where it suffers multiple scattering
processes, is detected by several microphones as it escapes through the
boundaries. These transducers can also act as loudspeakers and the
registered signal is played back in the time reversed sequence. Thus, the
signal focalizes in the source point forming a Loschmidt Echo \cite{Jalabert}.
According to the existing theory, an exact control of the wave function in
the cavity would require the control of the wave function and the normal
derivative at the boundaries. However, the reversal is quite good even when
these conditions are not fulfilled by the experiment: the detectors might
not enclose the cavity or the recording time period could be reduced to a
fraction. Another surprising feature of this time reversion procedure is
that it shows a much better stability in inhomogeneus media as compared to
ordered ones. This leads to numerous applications in medical physics 
\cite{Montaldo} and communications \cite{Edelman}. A first step to asses the
errors is to develop a procedure that could achieve perfect reversal. This
task was developed for the domain of quantum waves and was named Perfect
Inverse Filter (PIF) \cite{Pastawski}. The PIF procedure assures the exact
reversion by injecting a wave function that compensates precisely the
feedback effects through a frequency dependent renormalization that involves
the exact Green's function at the injection sites. Here, we use a simple
microscopic model that presents wave behavior and describes energy
dissipation to show that the PIF procedure also applies in the classical
domain. The model, represented in Fig. \ref{fig_scheme}, is a variation of
that of Rubin \cite{Rubin}: a surface oscillator with mass $m_{0}$ and
natural frequency $\omega _{0}$ is coupled to a semi-infinite harmonic chain
of bulk oscillators with mass $m$. 
We are interested in the time reversion of the initial condition where all 
the oscillators are in their equilibrium positions except for the surface 
one. The energy stored in the surface oscillator is expected to decay due 
to the effective friction produced by the \textquotedblleft environment%
\textquotedblright of light masses. The Hamiltonian is:
\begin{equation}
H=\sum_{i}\left( \frac{p_{i}^{2}}{2m_{i}}+\frac{m_{i}\omega _{i}^{2}}{2}%
u_{i}^{2}\right) +\sum_{i}\frac{K_{i,i+1}}{2}(u_{i}-u_{i+1})^{2},
\end{equation}
where $\omega _{i}$ is the natural frequency of the $i$th oscillator and $%
u_{i}$ describes the displacement from the equilibrium. Springs with elastic
constant $K_{i,i+1}$ accounts for the coupling between first neighbors. For
the proposed model, all springs and masses are equal and $\omega _{i}=0$,
except for the heavier surface mass that suffers an additional harmonic
restitutive force ($\omega _{0}$ is finite). The exchange frequency $\omega
_{\mathrm{x}}=\sqrt{K/m}$ and the ratio $\alpha =m/m_{0}<1$ have been chosen
to set the system in the extended-oscillatory dynamical regime \cite{Calvo}.
The equations of motion in the frequency domain can be written in the matrix
form%
\begin{equation}
\mathbb{D}^{-1}(\omega )\mathbf{u}(\omega )=\left( \omega ^{2}\mathbb{I}-%
\mathbb{M}\right) \mathbf{u}(\omega )=0, \label{eq_motion}
\end{equation}%
where $\mathbb{D}(\omega )=\left( \omega ^{2}\mathbb{I}-\mathbb{M}\right)
^{-1}$ is the resolvent associated to the dynamical matrix in the site basis%
\begin{equation}
\mathbb{M}=\left( 
\begin{array}{cccc}
\omega _{0}^{2}+\alpha \omega _{\mathrm{x}}^{2} & -\alpha \omega _{\mathrm{x}%
}^{2} & 0 & \cdots  \\ 
-\omega _{\mathrm{x}}^{2} & 2\omega _{\mathrm{x}}^{2} & -\omega _{\mathrm{x}%
}^{2} &  \\ 
0 & -\omega _{\mathrm{x}}^{2} & 2\omega _{\mathrm{x}}^{2} &  \\ 
\vdots  &  &  & \ddots 
\end{array}%
\right) .
\end{equation}%
The resolvent provides the solutions to Eq. \ref{eq_motion} with impulsive
forces $f_{i}(t)=m_{i}\Delta \dot{u}_{i}(0)\delta (t)$. This is,%
\begin{align}
u_{j}(t)& =\sum_{i}\int \frac{\mathrm{d}\omega }{2\pi }e^{-\mathrm{i}\omega
t}D_{j,i}(\omega )\Delta \dot{u}_{i}(0) \\
& =\sum_{i}D_{j,i}(t)\Delta \dot{u}_{i}(0).
\end{align}%
Notice that $D_{j,i}(t)$ relates the $j$th displacement amplitude due to an
initial condition of velocity in the $i$th oscillator. In general, the
solution of the Eq. \ref{eq_motion} in presence of forces $F_{i}(t)$ results%
\begin{equation}
u_{j}(t)=\sum_{i}\int_{0}^{t}\chi _{j,i}(t-t^{\prime })F_{i}(t^{\prime })%
\mathrm{d}t^{\prime },
\end{equation}%
and can be rewritten as%
\begin{equation*}
u_{j}(\omega )=\sum_{i}\mathbb{\chi }_{j,i}(\omega )F_{i}(\omega
)=\sum_{i}D_{j,i}(\omega )\left( \dfrac{-1}{m_{i}}\right) F_{i}(\omega ),
\end{equation*}%
where $F_{i}(\omega )=f_{i}(\omega )+g_{i}(\omega )$ is the Fourier
transform of the force applied at mass $i$ that is a sum of two components:
an impulsive force and\ a shifting force. This last is able to produce an
\textquotedblleft instantaneous\textquotedblright\ shift $\Delta u_{j}(0)$
in the position without changing its momentum. This would require that a
first \textquotedblleft impulsive kick\textquotedblright\ be followed by a
compensating one:%
\begin{equation*}
g_{i}(t)=\lim_{\tau \rightarrow 0}m_{i}\Delta u_{i}(0)\frac{1}{\tau }\left[
\delta \left( t+\tfrac{1}{2}\tau \right) -\delta \left( t-\tfrac{1}{2}\tau
\right) \right] ,
\end{equation*}%
which in frequency domain means:%
\begin{equation}
g_{i}(\omega )=-\mathrm{i}\omega m_{i}\Delta u_{i}(0).
\end{equation}%
Thus, in the time domain%
\begin{align}
u_{j}(t)& =\sum_{i}\int \frac{\mathrm{d}\omega }{2\pi }(-\mathrm{i}\omega
)\chi _{j,i}(\omega )e^{-\mathrm{i}\omega (t-t^{\prime })}m_{i}\Delta
u_{i}(t^{\prime }) \\
& =\sum_{i}G_{j,i}(t-t^{\prime })\Delta u_{i}(t^{\prime }),
\end{align}%
which serves as a definition for the position-position response function,
also called the Green's function:%
\begin{equation}
G_{j,i}(\omega )=-\mathrm{i}\omega D_{j,i}(\omega )=\mathrm{i}\omega \chi
_{j,i}(\omega )m_{i}.
\end{equation}%
On the other hand we will use linearity to write the observed displacement
in terms of $\delta u_{i}(t)$, the forced position shift accumulated in the
unit time:%
\begin{equation}
u_{j}(t)=\sum_{i}\int G_{j,i}(t-t^{\prime })\delta u_{i}(t^{\prime })\mathrm{%
d}t^{\prime }.
\end{equation}%
We seek the injection function $\delta u_{i}(t)$ that produces the exact
reversion of the original wave within the control region, i.e. $%
u_{j}^{\mathrm{rev}}(t)\equiv u_{j}^{\ast }(2t_{R}-t)$ for $t_{R}\leq t\leq
2t_{R}$. According to Ref. \cite{Pastawski}, the perfect time reversal is
possible if the dynamics starts and ends up without any excitation inside
the cavity. Even when our system starts with an \textquotedblleft
excited\textquotedblright cavity, the lack of momentum at each mass ensures
that forward and backwards evolutions are identical. Once the decay signal is
registered at the transducer for positive time, the earlier time values 
(corresponding to a fictitious injection) are also known and we build the 
function to be inverted at time $t_{R}$ 
\begin{equation}
\tilde{u}_{s}(t)=\left\{ 
\begin{array}{cc}
u_{s}^{\ast }(t_{R}-t), & -t_{R}\leq t\leq 0 \\ 
u_{s}(t-t_{R}), & 0\leq t\leq t_{R}%
\end{array}%
\right. ,
\end{equation}%
and the injection function can be obtained in the frequency domain by%
\begin{equation}
\delta u_{s}(\omega )=\frac{\tilde{u}_{s}^{\mathrm{\ast }}(\omega )}{%
G_{s,s}(\omega )}.  \label{eq_injPIF}
\end{equation}%
This equation defines the Perfect Inverse Filter (PIF) for a classical wave,
where the injection prescribed by the TRM procedure appears now corrected by
the Green's function $G_{s,s}(t)$.
The time evolution of the displacement amplitude is shown in Fig. \ref%
{fig_dynamic} for several situations. The recording time $t_{R}$ is longer
than the decay time in the whole cavity, i.e. all the masses in the control
region have enough time to recover their equilibrium positions.
\begin{figure}[ptb]
\begin{center}
\includegraphics*[width=7.5cm]{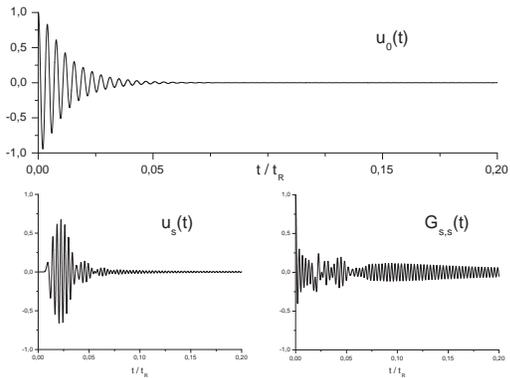}
\end{center}
\caption{Displacement amplitudes. Up: initial condition at the surface
oscillator. Down - left: registered signal in $x_{s}$. Down - right:
response function in $x_{s}$. The chosen parameters are: $\protect\alpha%
=0.25 $, $\protect\omega_{0}$ $=$ $1.5\protect\omega_{\mathrm{x}}$, $%
t_{R}=1000\protect\omega_{\mathrm{x}}^{-1}$ and $x_{s}=10$.}
\label{fig_dynamic}
\end{figure}
The injection functions of both procedures differs near the bandedges ($%
\omega =\pm 2\omega _{\mathrm{x}}$) indicating that TRM corrections are
importants in cases where the whole spectrum is involved (typically in
broadband experiments). When all the masses in the cavity are in their
equilibrium position, the injection at the source point $x_{s}$ produces
oscillations that propagate both sides of $x_{s}.$ Only dynamics inside the
cavity is fully reversed. Fig. \ref{fig_decrec} shows the local energy of
the surface oscillator 
\begin{equation*}
E_{0}(t)=\tfrac{1}{2}m_{0}\left\vert \dot{u}_{0}(t)\right\vert ^{2}+\tfrac{1%
}{2}m_{0}\left( \omega _{0}^{2}+\alpha \omega _{\mathrm{x}}^{2}\right)
\left\vert u_{0}(t)\right\vert ^{2};
\end{equation*}%
its decay and recovering presents the three temporal domains \cite{Rufeil}:
It begins with a quadratic dependence, continues with an exponential decay
associated with a Self Consistent Fermi Golden Rule, and becomes an inverse
power law when the energy return is comparable to the residual energy in the
surface mode. The time reversed signal reproduces all the three regimes,
even when they involve very different signal intensities.
We attempt to asses the quality of the reversal by the Loschmidt Echo \cite%
{Jalabert} of a wave function $\varphi (t)$ normalized in the cavity:%
\begin{equation}
M(t)=\left\vert \left\langle \varphi _{\mathrm{rev}}(t)\right\vert \left.
\varphi (2t_{R}-t)\right\rangle \right\vert ^{2}\text{ \ \ }t_{R}\leq t\leq
2t_{R}.
\end{equation}%
First, the inner product uses a metric tensor given by the Hamiltonian, i.e.
we refer the recovered energy to that originally contained within the
control region. Considering the case where $t_{R}=1000\omega _{\mathrm{x}%
}^{-1}$, the PIF yields $M_{\mathrm{PIF}}=0.999$ while the TRM gives $M_{%
\mathrm{TRM}}=0.982$. This result is not representative of the evidenced by
Fig. \ref{fig_recovering}. Alternatively, we use the Euclidean metric
tensor, obtaining $M_{\mathrm{PIF}}=1$ and $M_{\mathrm{TRM}}=0.765$, i.e. in
PIF all the initial condition have been recovered whereas in TRM there is a
spreading of displacements and velocities along the cavity.
The reversed local energy in both procedures is compared in Fig. \ref%
{fig_energy}. The PIF procedure can not be distinguished from the ideal
reversal. The TRM has two failures: the amplitude of the local energy at $%
t=2t_{R}$ is always less than the initial case and the temporal regimes are
delayed.
\begin{figure}
\begin{center}
\includegraphics*[width=7.5cm]{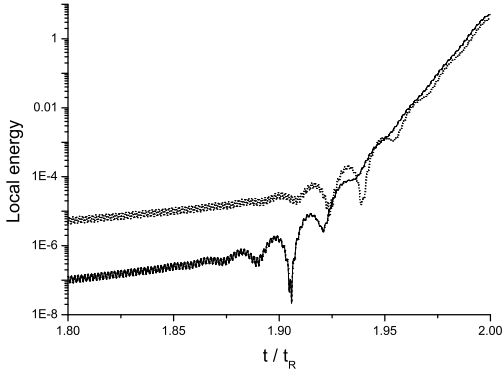}
\end{center}
\caption{Focalization of the local energy in logaritmic scale. Solid line
represents the ideal time reversed decay, dashed line is PIF and TRM is
shown in dotted line. Note that PIF is superposed with the ideal case.}
\label{fig_energy}
\end{figure}
In summary, the time reversion of the dynamics of a pendulum coupled to an
harmonic chain has been done by means of two types of \textquotedblleft
stubborn daemons\textquotedblright : 1) The TRM, which neglects memory
effects (meaning that an instantaneous system response is assummed). 2) The
PIF, that accounts for memory and feedback. This last is much better if the
initial condition is build up over the whole range of frequencies. In this
case, the correction of the Green's function becomes non trivial and ensures
a better reversion quality.
\appendix
\section{Classical dynamics through pair partitioning}
While the 1D dynamics in the model considered can be obtained analytically
by the continued fraction method \cite{Medina}, we also evaluate it
numerically by developing an algorithm, the pair partitioning, inspired in
the Trotter strategy used in quantum dynamics \cite{De Raedt}. We split the
kinetic terms to rewrite the Hamiltonian as%
\begin{equation}
H=\sum_{i}H_{i,i+1}.  \label{eq_trotter}
\end{equation}%
Now, each term represents an effective Hamiltonian for two coupled
oscillators, with twice the mass and half of the natural frequency each.
Pair dynamics is solved analytically and impose a periodic evolution
sequence that alternates each coupled pair according to their parity. The
total energy is not exactly conserved but fluctuates with an amplitude $%
\Delta E$ around the ideal conserved value. Since $\Delta E$ is proportional
to $\delta t^{2}$, the square of the temporal step, it becomes negligible
for typical cases where $\delta t=0.01\omega _{\mathrm{x}}^{-1}$. The fact
that each piecelike dynamics is perfectly reversible is very important for
the test of different time reversal procedures.

\end{document}